\documentclass[twocolumn,prb,aps]{revtex4}
\usepackage{graphicx}
\newcommand{\beq}{\begin{eqnarray}}
\newcommand{\eeq}{\end{eqnarray}}
\begin{document}
\draft
%\twocolumn
%\preprint{dvi file made on \today}
%\input epsf.sty
%\input psfig.sty

\title
{Minimal Model for Disorder-induced Missing Moment of Inertia in Solid
 $^4$He}

\author{Jiansheng Wu and Philip Phillips }
\affiliation{Loomis Laboratory of Physics,
University of Illinois at Urbana-Champaign,
1100 W.Green St., Urbana, IL., 61801-3080}

\begin{abstract}
The absence of a missing moment inertia in clean solid $^4$He
suggests that the minimal experimentally relevant model is one in
which disorder induces superfluidity in a bosonic lattice.  To this
end, we explore the relevance of the disordered Bose-Hubbard model in
this context.  We posit that a clean array $^4$He atoms is a
self-generated Mott insulator, that is, the $^4$He atoms constitute
the lattice as well as the `charge carriers'. With this assumption,
we are able to interpret the textbook defect-driven supersolids as
excitations of either the lower or upper Hubbard bands.  In the
experiments at hand, disorder induces a closing of the Mott gap
through the generation of mid-gap localized states at the chemical
potential. 
Depending on the magnitude of the disorder, we find that the
destruction of the Mott state takes place for $d+z>4$ either
through a Bose glass phase (strong disorder) or through a direct
transition to a superfluid (weak disorder). For $d+z<4$, disorder is always relevant. The critical value of
the disorder that separates these two regimes is shown to be
a function of the boson filling, interaction and the momentum cut
off. We apply our work to the experimentally observed
  enhancement $^3$He impurities has on the onset temperature for the missing moment of inertia.
   We find quantitative agreement with experimental trends.
\end{abstract}
\maketitle

\section{Introduction}

While superflow in a state of matter possessing a shear
modulus might initially seem untenable,
experimental claims for precisely this phenomenon
in solid $^4$He now abound\cite{vycor,KC,impeff,Chan06,null1,null2,null3,SRR,rittner,SB,g1,g2,science}.   Reported in the experiments
by Kim and Chan\cite{vycor,KC} (KC) was a dramatic change below $200mK$ in the period
of a torsional oscillator containing solid $^4$He.  Because
superfluids come out of equilibrium and detach from the walls of the
rotated container, they  are expected to give rise to a period shift
in such a geometry, assuming, of course, the rotation velocity is
less than the critical velocity to create a vortex. The result is a
missing moment of inertia\cite{leggett,chester67} (MMI) and hence the period
of oscillation decreases.  The magnitude of the MMI is a direct
measure of the superfluid fraction.  In the original experiments reported by Kim and Chan\cite{vycor,KC},
the superfluid fraction ranged from .14\% for $^4$He in vycor\cite{vycor} to 2\% in bulk $^4$He.  However, Rittner and Reppy\cite{SRR,rittner} have shown that the quench time for solidifying the liquid is pivotal in determining the superfluid fraction.  In one extreme, when the sample is fully annealed, no MMI occurs.   In the other, the MMI increased to an astounding 20\% in samples in which the solidification from the liquid occurred in less than 2 minutes. While not all groups\cite{science} have been able to eliminate the MMI signal entirely by annealing\cite{null1,null2,null3} the sample and in fact there is at least one claim of MMI in a single crystal\cite{channew}, the enhancement of MMI by a rapid quench does not seem to be in question.  In fact, two independent experiments point to the key role played by disorder: 1) the Todoschenko et. al.\cite{todo} measurement that the melting curve of $^4$He remained unchanged from the $T^4$ law expected for phonons in ultra-pure samples with a $^3$He concentration of 0.3ppb and 2) the experiments of Clark and Chan\cite{impeff} that increasing the $^3$He
impurity concentration\cite{impeff} from 20ppm to 40ppm increases the
transition temperature from 0.35K to 0.55K.

Clearly the standard textbook supersolid in which vacancy or
interstitial defects Bose condense\cite{AL,anderson} fails to
explain the disorder dependence of the MMI. 
  In fact, it is unclear at this writing even if a super-component is needed\cite{avb} to explain the MMI,
  primarily because experiments\cite{beamish} designed to detect persistent mass flow have revealed no telltale signature.  Monte Carlo simulations\cite{gb1} reveal, however, that superflow in solid $^4$He is confined to grain boundaries.  This observation is supported by the experiments of Sasaki et al.\cite{SB} who observed
       mass flow only in samples containing grain boundaries. Nonetheless, the precise
      relationship between this experiment and
      the torsional oscillator measurements is unclear because
       mass flow was observed at
      temperatures (1.1K which is not far from the bulk
       superfluid transition temperature) vastly exceeding
       the onset temperature for MMI in the torsional oscillator experiments\cite{KC}, namely $T_c=0.2K$.

Even if the MMI is not tied to superflow, disorder is still the key
player underlying the experimental observations\cite{science}. As disorder can
 induce superfluidity in the disordered Bose-Hubbard model,
we explore its utility as a minimal model for the experimental observations.
Certainly, this model does not have all of the microscopic
details necessary to describe $^4$He, in particular the precise details needed to describe a grain boundary or the long-range interactions between $^4$He atoms.  Our central claim is only that it serves as a minimal
model to describe disorder-induced superflow in a bosonic system. Our
work is based on a simple claim: $^4$He is a hexagonally
close-packed self-generated Mott insulator.  In a self-generated or self-assembled Mott insulator, the
lattice and the `charge carriers' are one and the same. In contrast, in Fermionic Mott insulators, the electrons occupy pre-existing lattice sites formed by the ions.  Our characterization of $^4$He as a self-generated Mott insulator is relevant for three reasons: 1) In a supersolid the relevant transport is of the $^4$He atoms themselves. Hence, if they form a Mott insulator in the clean system, no transport is possible. 2) Experiments\cite{science} and simulations find an
absence of MMI in the clean limit\cite{gb1,CC}. 3) We can immediately classify the candidate supersolids with this scheme because disorder can
either\cite{huse} 1) self-dope the system\cite{Sawatzky,vac} or 2)
create mid-gap states\cite{fisher}.   The former would generate
either vacancies or interstitials and hence excitations in
either the lower or upper Hubbard bands. The Andreev/Lifshitz\cite{AL} scenario in which vacancies or interstitials Bose condense can be thought of as arising from doping a self-generated Mott insulator.
We call such a state SS1.  In electronic systems, disorder is
well-known to have such an effect\cite{Sawatzky}.  We will show that
SS1 does not obtain in the disordered Bose Hubbard model.  Rather a
superfluid state (SS2) forms from overlapping localized mid-gap
 states\cite{fisher,scall,bh1,bh2,bh3,bh4,bh5,herbut,bh7}.
 We argue that SS2 is most relevant to the experimental observations.
 
We establish several new results in this paper.  First, we use the
replica technique coupled with a renormalisation group analysis to
show that 
weak and large disorder disrupt the Mott insulator (MI) in radically different ways.
 In particular, the critical value of the disorder that separates these two regimes is
 a decreasing function of filling.   Second, in the weak disorder regime a direct transition from
 the superfluid (SF) to the Mott insulator is possible whereas such a transition always involves
 the Bose glass (BG) phase at large disorder. This result resolves the controversy\cite{bh3,bh5,herbut}
 surrounding when the destruction of the superfluid necessitates an intermediate Bose glass phase.
 Finally, we offer a quantitative test of this model by applying it to the $^3$He enhancement of $T_c$.
 The  quantitative agreement suggests that the essence of the MMI in the experiments is
 captured by the disordered-Bose Hubbard model.

\section{Initial considerations}

To describe boson motion in a random potential, we adopt the site-disordered Bose-Hubbard model
\beq
 H=- t \sum_{\langle i,j\rangle} \left(b_i^{\dag} b_j + c.c \right)
+\sum_{i} \epsilon_i n_i+\frac{V}{2} \sum_{i} n_i(n_i-1) .\nonumber\\
\eeq
 In this model,
$b_i^{\dagger}$ is the creation operator for a boson at site $i$ and
$n_i$ is the particle number operator and $t$  and $V$ are the
Josephson coupling and on-site repulsion, respectively. We also
define a value $J=z t$ where $z$ is the number of nearest
neighboring sites.

Though much of the theoretical work\cite{scall,bh1,bh2,bh4,herbut,bh7} on the disordered Bose-Hubbard model
has confirmed the originally proposed picture that an intermediate Bose glass localized phase disrupts the MI-SF transition,
several key issues remain.

P1.\quad {\it Is there a direct MI-SF transition in the presence of disorder?}

For example, several analytical treatments\cite{fisher,herbut,bh7,bh4} suggest
that the  Bose-glass phase completely surrounds the Mott insulating phase, making a direct transition from the MI to SF impossible.
 However, simulations\cite{scall,bh1,bh3} and a renormalization group analysis\cite{bh5}
 find that a Bose glass is absent in $d=2$ at commensurate fillings.
  In fact, the renormalization group analysis of Pazmandi and
  Zimanyi\cite{bh5}
  lays plain that the weak and strong disorder cases are fundamentally different.
 Only in the strong disordered case does the Bose glass phase completely surround the Mott lobes.
 However, Herbut\cite{herbut} has also provided a convincing treatment of the large-filling limit and
 concluded that disorder is always relevant and destruction of the superfluid obtains through the Bose glass
 even in $d=2$. 

P2.\quad {\it Do Mott insulators vanish for unbounded distributions?}

Fisher, et al.\cite{fisher} argued that no Mott insulating phases are possible when  the width of the disorder exceeded $V/2$ at $T=0$.  Consequently, for unbounded distributions, Mott insulators are absent at $T=0$ and only a superfluid phase exists\cite{fisher}. Does the same hold for finite temperature?  As the distributions characterizing disorder\cite{demarco} in optical lattices are typically unbounded, this question must be resolved.

\subsection{Resolution}

We resolve both of the problems in this paper.

First, we show that the missing ingredients that squares these seemingly
contradictory results in P1 are 1) dimensionality, 2) critical momentum cutoff
$\Lambda_c$ and 3) a filling and interaction-dependent
critical value of the disorder $\Delta_c$.
For $\epsilon=4-(d+z)>0$,
disorder is always relevant. In this case, the Mott insulating phase is destroyed and a BG obtains. This is in agreement with the work of Herbut\cite{herbut} on the destruction of superfluidity in $d=1$ and $2$ always takes place through the Bose glass.  He finds that $z=1.93$, implying that $\epsilon>0$, matching our criterion for the relevance of disorder.

\begin{figure}
  %  \centering
        \includegraphics[scale=0.60]{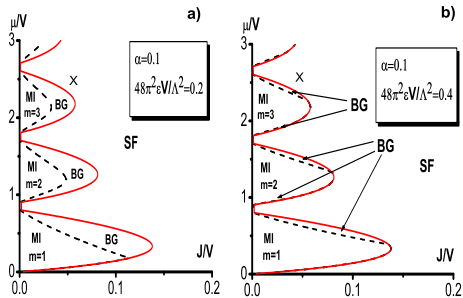}
    \caption{Phase diagram for the disordered Bose-Hubbard model as a function of chemical potential $\mu/V$ and hopping strength, $J/V$.
    MI,BG and SF stands for Mott insulator, Bose glass and superfluid respectively. In the present of disorder, the lobes are shrunk,
    and we have two phase insider the lobes,
    MI and BG and outside, we have SF phase. a) The typical phase diagram when $\epsilon<0$
    and $\Lambda>\Lambda_c$ or when $\epsilon<0$,
    $\Lambda<\Lambda_c$ and $\Delta>\Delta_c$. In this case, direct transition from MI to SF is only possible at $m=1$. b) The typical phase
    diagram when $\epsilon<0$,$\Lambda<\Lambda_c$ and
    $\Delta<\Delta_c$. In this case, direct transition from MI to SF
    is possible for many filling numbers. }
    \label{bg}
\end{figure}

For systems with $\epsilon<0$, there exists a boundary in phase
space separating disorder relevant and disorder irrelevant regions.
For filling $m=1$, a direct transition is always allowed. For large
fillings, the situation is more complicated. If the momentum cutoff
(determined by the lattice constant) exceeds a critical value,
$\Lambda>\Lambda_c$, the MI is surrounded by a BG  phase and direct
transition from MI to SF is forbidden as illustrated in Fig.
(\ref{bg}). In the opposite regime, $\Lambda<\Lambda_c$, the
strength of the disorder is the key ingredient. For the weak
disorder case, $\Delta<\Delta_c$, a direct transition is allowed for
large fillings while it is forbidden for strong disorder,
$\Delta>\Delta_c$. These results are in accord with the RG analysis
of Pazmandi and Zimanyi\cite{bh5} who studied an infinite range
model and found\cite{bh5} that for $\epsilon<0$ a direct
transition is possible.  For $\epsilon>0$, they found that disorder
is in general relevant except at perhaps the particle-hole symmetric
point at small filling where a direct transition survives at weak
disorder.

%From our
%replica analysis of the Gaussian distribution, if we define a
%critical disorder $\Delta_c$ as Eq.().

 %%Here $\Lambda=\pi/a_0$ is the momentum cut off and $a_0$ is the
%%lattice constant. We show that once the disorder lies outside this
%%domain, the Bose glass always surrounds the  Mott lobe for large
%%filling number, thereby preventing a direct transition between the
%%Mott insulator and superfluid phases for large filling.

%Since $\Delta_c$
%depends on the filling, Eq. (\ref{deltac}) cannot be satisfied
%simultaneously for all integer fillings of the lobes. Consequently,
%the phase that obtains once the Mott insulator is destroyed can vary
%with filling as illustrated in Fig. (\ref{bgphase}).
%In the large
%filling limit, $m\rightarrow\infty$, $\Delta_c$ vanishes, and we
%recover the result of Herbut\cite{herbut} that disorder is always
%relevant and a Bose glass necessarily results from the destruction
%of the Mott insulator. This appears to be the first time such a
%filling-dependent criterion has been derived.

In addition,  we analyse a Gaussian distribution for the site energies here and demonstrate how temperature and disorder are intertwined.  At finite temperature, we establish the existence of integer-filling Mott states. However, the $T=0$ analysis is beyond the scope of the treatment here as it corresponds to the infinite disorder limit.  In particular, our replica analysis on unbounded distributions is valid strictly when 
\beq
\beta\Delta^2/V<1
\eeq
where $\Delta$ is the variance of the distribution and $V$ the on-site repulsion.  In fact, this breakdown is fundamentally related to our central point that for bosonic systems, disorder destroys Mott insulators and gives rise to superfluids. To see how this comes about, it is sufficient to integrate out the randomness by
using the replica trick,
\beq \ln
Z=\lim_{n\rightarrow 0}\frac{Z^n-1}{n},
\eeq
 in which $n$ represents
the number of replicas and $Z$ is the partition function. Performing
the integral over the disorder
\beq
  Z^n  &=& \prod_i \int d \epsilon_i \int D b_i D b_i^{\dag} e^{-\frac{(\epsilon_i-(-\mu))^2}{2\Delta^2}} e^{-\beta\sum_a H^a}\nonumber\\
    &=&  \prod_i\int D b_i D b_i^{\dag} e^{-\beta H_{\rm eff}}\nonumber\\
    H_{\rm eff}&=& - t \sum_{\langle i,j>,a} \left(b_{i}^{a\dag} b^a_{j} + c.c \right) -\sum_{i,a} (\mu+V/2) n^a_{i}\\&+& \frac{V-\beta \Delta^2}{2}
     \sum_{i} (n^a_{i})^2- \sum_{i,a\neq b}\frac{\beta \Delta^2}{2} n^a_{i}n^b_{i}\nonumber
\eeq
results in an effective Hamiltonian for the disordered problem.
Here $a$ is the replica index and we have assumed that the disorder
is described by a Gaussian distribution of width $\Delta$.  We see clearly that the
on-site interaction is replaced by
\begin{eqnarray}\label{VC}
    V_{\rm  eff}&=& V-\beta \Delta^2.
\end{eqnarray}
Consequently, at sufficiently low temperature, disorder can destroy the Mott gap.

This paper is organized as follows. In the next section we compute the phase diagram for the disordered Bose Hubbard model using replicas and a renormalization group analysis. Since we start our analysis from the strongly interacting regime, any diagrams that are calculated cannot be computed using Wick's theorem.  To circumvent this problem, we resorted to the analysis detailed in the Appendix. We explicitly compare the results for Gaussian and the uniform distribution case studied earlier\cite{fisher}.  The analysis of the phase boundary for the Bose glass is presented at the end of this section.  This analysis is particularly lengthy as the topology of the phase boundaries is found to be delicately determined by the strength of the disorder and the cutoff.  We close with an application of our central result that disorder enhances superfluidity to the problem of $^3$He-induced enhancement of the onset temperature for missing moment of inertia.

%===============================================================
% First Step: MI and SF
%===============================================================

\section{Phase Diagram of the Disordered Bose Hubbard Model}

In this section, we derive the phase diagram for the disordered
Bose-Hubbard model for the Gaussian and uniform distribution of site
energies.  To establish the phase boundaries for the Mott insulator
(MI) and superfluid (SF) phases, we employ a saddle-point analysis
on the partition function\cite{fisher,RSY,WP,DP,DP2,DP3},
\begin{eqnarray}\label{Z}
Z&=& Z_0 \int \prod_i D \psi_i(\tau)D \psi^*_i(\tau)\exp[-S(\psi_i)]\\
S(\psi)&=&
\sum_{i,j}[J^{-1}]_{ij}\psi_i^*(\tau)\psi_j(\tau)\nonumber\\
&-&\sum_i\ln\left\langle T_{\tau}\exp\left[\int \tau
\psi_i(\tau){b_i^a}^{\dag}+H.c.\right]\right\rangle_0
\end{eqnarray}
by introducing a Hubbard-Stratonovich field $\psi_j$ to release the $b_i^{\dag}b_j$ term. Appearing in Eq. (\ref{Z}) are $[J^{-1}]_{ij}$, the inverse matrix of hopping rates which
will determine the band structure for the kinetic energy and
$Z_0=Tr\exp(-\beta H_0)$,
$b_i^a(\tau)=e^{H_0\tau}b_i^a(0)e^{-H_0\tau}$.

Differentiating the free energy  with respect to $\psi$ yields the saddle
point equation\cite{PPbook}
\begin{eqnarray}
     \sum_j [J^{-1}]_{ij}\psi^a_i(\tau) &=& \left\langle
b_{i}^a(\tau) \right\rangle.
%     \frac{1}{J}&=& \int_0^{\beta}d \tau \int_0^{\beta} d\tau' \left\langle T S_{\mu}^a (\tau) S_{\nu}^a (\tau')\right\rangle.
\end{eqnarray}
Because $\psi_i^a$ is linearly related to $\left\langle b_i^a\right\rangle$,
its average value will serve to define the superfluid order parameter.
This can be seen more clearly by performing the cumulant expansion on ${b_i^a}^{\dag}(\tau)$.   The action can then be rewritten as,
\begin{eqnarray}\label{action}
 S(\psi)&=& \beta \left[\sum_{i,a} r_{ij}\psi_i^{a*}\psi_j^a+c.c+u\sum_{i,a} |\psi_i^a|^4\right. \nonumber\\
 &+& \left.v\sum_{i,a\ne b}|\psi_i^a|^2|\psi_i^b|^2+O(|\psi|^6)\right]\nonumber\\
 r_{ij}= [J^{-1}]_{ij}&-&\delta_{ij}\int_0^{\beta} \int_0^{\beta} d\tau d\tau' \left\langle T {b_{i}^a}^{\dag} (\tau) b_{i}^a
 (\tau')\right\rangle \label{r}
\end{eqnarray}
where $r$ matrix acts as the mass term and hence determines the appearance of superfluid phase.

\subsection{ Gaussian disorder}

For the Gaussian case, the Hamiltonian consist of two parts,
\begin{eqnarray}\label{H}
    H_0&=& \frac{V_{\rm eff}}{2} \sum_{a,i}(n_i^a)^2- \frac{\beta \Delta^2}{2}\sum_{i,a\neq b} n_i^a n_i^b-\mu_{\rm eff}\sum_{a,i}n_i^a \nonumber\\
     H_1&=& -t\sum_{\langle i,j\rangle} {b_i^a}^{\dag}  b_j^a+c.c,
\end{eqnarray}
where $\mu_{\rm eff}=\mu+V/2$. Because the hopping term is a perturbation,
our theory is valid strictly for $V>J$.  In addition, since we are working in the limit in  which the Mott lobes are well-formed, we must have that $V_{\rm eff}>0$ and $\beta V\gg 1$.  The latter two constraints can be written as
$1>\alpha$ where $\alpha=\beta\Delta^2/V$. It is this parameter that we will use to characterize the strength of the disorder.
Using the eigenstates of $H_0$, that is, the eigenstates of
particle number, $\left\langle  m| \theta  \right\rangle =\frac{1}{2\pi} e^{i
\sum_a m^a\theta }$, we have,
\begin{eqnarray}
&& \left\langle  T b_i^{a\dag} (\tau) b_{j}^b
(\tau')\right\rangle_0=\\
&\times &\frac{1}{Z_0}\sum_{m}\left[ \left\langle m|e^{H_0\tau}{b_i^a}^{\dag}e^{-H_0\tau}e^{H_0\tau'}{b_j^b} e^{-H_0\tau'}|m\right\rangle\theta(\tau-\tau')\right.\nonumber\\
&+&\left.
\left\langle m|e^{H_0\tau'}{b_j^b}e^{-H_0\tau'}e^{H_0\tau}{b_i^a}^{\dag}e^{-H_0\tau}|m\right\rangle\theta(\tau'-\tau)\right].
\end{eqnarray}
For the above to be nonzero, we have to choose $a=b$ and $i=j$.
Inserting a complete set of states, $1=\prod_c \sum_{m^c=1}^{\infty}|m^c \rangle\langle
m^c|$, between $b_i^{a\dag}$ and $b_i^a$, we have only two terms left,
$|m_a\pm 1 (c=a) m_c (c\ne a) \rangle\langle m_a\pm 1 (c=a) m_c (c\ne a)|$.
Note that we have replica symmetry between the initial and final
states. However,  replica symmetry breaking must be present in the
intermediate states to have a nonzero correlation. The inserted state
together with the creation and annihilation operators will lead to a
term of the form $E_0(m_i^a\pm 1,m_i^b)-E_0(m_i^a,m_i^b)$ where
$E_0(m_i^a,m_i^b)$ is the eigenenergy of $H_0$. The explicit form for this term is
\begin{eqnarray}\label{}
E_0(m_i^a,m_i^b)&=&\frac{V_{\rm eff}}{2} \sum_{a,i}(m_i^a)^2-
\frac{\beta \Delta^2}{2}\sum_{i,a\neq b} m_i^a
m_i^b\nonumber\\
&-&\mu_{\rm eff}\sum_{a,i}m_i^a.
\end{eqnarray}

After integrating over $\tau$ and $\tau'$, we obtain,
\begin{eqnarray}\label{pb}
&&\int d\tau\int d\tau' \left\langle  T b_i^{a\dag} (\tau) b_{j}^b
(\tau')\right\rangle_0 \nonumber\\
&=&\frac{(m+1)}{\varepsilon_+} \left(1-\frac{1}{\beta \varepsilon_+}
\right)+
\frac{m}{\varepsilon_-} \left( 1-\frac{1}{\beta \varepsilon_-} \right)\nonumber\\
&\approx & \frac{(m+1)}{\varepsilon_+} +\frac{m}{\varepsilon_- },
\end{eqnarray}
where we can neglect $1/\beta\varepsilon_{\pm}$ only when the
temperature is small relative to the Mott gap, that is, $k_B
T/\varepsilon_{\pm}\ll 1$. In the above equation, we are considering
the energy of one replica, so the $m_i^a m_i^b$ term will give rise
to $(n-1)m_i^2$, part of which is linear in $n$. Because we will
take the limit $n\rightarrow 0$ in the end, we can neglect all the
high order terms when we calculate the energy of one replica. In
terms of $D=\beta\Delta^2/2$, the energies $\varepsilon_{\pm}$ are
defined as
\begin{eqnarray}\label{EGD}
\varepsilon_{\pm}&=& E_0(m_i^a\pm 1,m_i^b)-E_0(m_i^a,m_i^b)\\
&=& \frac{V_{\rm eff} }{2} \pm m \left(\frac{V_{\rm eff}}{2}-(n-1) \frac{\beta \Delta^2}{2} \right) \mp \mu_{\rm eff}\\
&=&\left\{  \begin{array}{l} \varepsilon_+(m)= m V-\mu-(m+1)D \nonumber\\
  \ \varepsilon_-(m)= (1-m)V+\mu+(m-1)D\  (\rm Gaussian) \end{array}\right.,
\end{eqnarray}
 We defined $m$ to be the integer
closest to $\mu_{\rm eff}/V_{\rm eff}$ because in the low
temperature limit, only this term in $Z_0$ dominates. This holds for a
system with non-conserved or
commensurate particle number. For a system with conserved and
incommensurate particle number, we should replace $m$ by the
particle number $m_i$ on each site.

For the single-component case, $r$ is a scalar and
we just need $r<0$ to have superfluid order. In our case, however,
 $r$ is a matrix which must be diagonalized. For
simplicity, we consider only nearest neighbor hopping in one
dimension case where $J_{ij}=t(\delta_{i,j+1}+\delta_{i,j-1})$. The
diagonal hopping matrix will be $ \delta_{ij}J\cos(\frac{2
j\pi}{N})$ with $j=0,1,...,N-1$ which is the quantum number of
momentum $k=2\pi n/L$. Here $N$ is the number of sites, $L$ the
system size, and $J=z t$ where $z$ is the number of nearest
neighbors. Diagonalizing the hopping matrix will of course require
various linear combinations of the $\psi_i$ fields. Such linear
combinations will leave $\left\langle  T b_i^{a\dag} (\tau) b_{i}^a
(\tau')\right\rangle_0 $ invariant because of the $\delta_{ij}$
appearing in front. Consequently, the condition for superfluid order
is,
\begin{eqnarray}
r_{ij}(n) \equiv \frac{1}{J\cos(2n\pi /N)}-\left\langle  T b_i^{a\dag}
(\tau) b_{i}^a (\tau')\right\rangle_0 &\le&  0.\nonumber
\end{eqnarray}
Note superfluid order arises anytime one of the $r_{ij}'s<0$.  The phase diagrams we construct in this section correspond strictly to phases in which $\psi_i=0$ and $\psi_i\ne 0$. In the Bose glass section, we will make the distinction between the localized phase being gapped or ungapped.

\begin{figure}
  \includegraphics[scale=0.65]{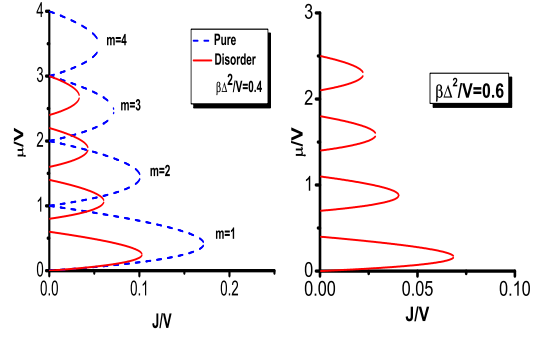}\\
  \caption{ Phase diagram for the disordered Bose-Hubbard model with a Gaussian distribution of site energies.  The two values of the disorder correspond to
 $\alpha=\beta\Delta^2/V=0.4$ and $\alpha=0.6$.}\label{fig2}
\end{figure}

The onset of a MI state is determined by the largest eigenvalue of $[J^{-1}]$. For a continuous band, this corresponds to $1/J$. Consequently, the phase
boundary separating MI and SF phases is given by
\begin{eqnarray}\label{SPE}
\frac{1}{ J}&=& \frac{(m+1)}{\varepsilon_+} +\frac{m}{\varepsilon_-
}
\end{eqnarray}
Using Eq. (\ref{EGD}), we rewrite Eq.(\ref{SPE}) as
\begin{eqnarray}
     m(m-1)V^2+V\left[(1-2m)\mu+J+D(m+1-2m^2)\right]&+&\nonumber\\
    \mu^2+\mu [J+2 m D]-(m+1)DJ+(m^2-1)D^2 =&0&.\nonumber
\end{eqnarray}
This equation describes a set
of super-planes in terms of $V-\Delta-\mu$  for different $m$. For a
given chemical potential,  it describes the phase boundary as a
function of disorder and $V$.
%================================================================
For $m=1$, that is, one boson per site, we recover exactly
$V_c(\Delta)$ (Eq. (\ref{VCD})) as the phase line between SF and MI.
%================================================================
The analogous expressions can also be derived for fixed disorder
$\alpha=D/V$ but varying chemical potential $y=\mu/V$ and $x=J/V$
which reads,
\begin{eqnarray}
%\frac{\mu}{V}&=& m-\frac{1}{2}+\frac{1+2mD}{2}\frac{J}{V}\pm\left[ 1-\left(4m+2+4D\right)\frac{J}{V}\right.\nonumber\\
%&+&\left.\left(1-(8m+4)D+4 D^2\right)\left(\frac{J}{V}\right)^2
%\right]^{1/2}
y&=& m-\frac{1}{2}-\frac{x}{2}-m \alpha\nonumber\\
&\pm& \frac{1}{2}\sqrt{(1-2\alpha)^2+\left(4m+2\right)(2
\alpha-1)x+x^2}.
\end{eqnarray}
The result is shown in Fig. (\ref{fig2}).  From the
figures, we see that the distance between the upper and lower boundaries
 of each lobe have shrunk by $2\alpha$ and the whole
lobe is shifted downward by $m\alpha$ relative to the ordered solution.  As is evident, the MI phase still exists at finite temperature for the unbounded distribution. Finally, increasing disorder decreases the size of the Mott lobes.
That the size of the Mott lobes shrinks with disorder has also been found in the extensive simulations of Trivedi and colleagues\cite{bh1} for a uniform distribution of site energies.

%$\mu<0$ region which means that MI of varying boson number cannot be
%accessed at zero temperature without changing the chemical
%potential. So the self doping is a statistical effect.

\subsection{ Uniform Distribution}

For completeness, we also compute the uniform
distribution of site energies of width $2\Delta$ studied in the original treatment of the disordered Bose-Hubbard problem\cite{fisher}. Integrating over the disorder in this case
is also straightforward and yields
\begin{eqnarray}
  Z^n  &=& \prod_i \int_{\Delta}^{\Delta} d \epsilon_i \frac{1}{2\Delta} \int D b_i D b_i^{\dag}  e^{-\beta\sum_a H^a} \nonumber\\
    &=&  \prod_i\int D b_i D b_i^{\dag} e^{-\beta H_{\rm eff}}\nonumber\\
    H_{\rm eff}&=& - t \sum_{\langle i,j\rangle,a} \left(b_{i}^{a\dag} b^a_{j} + c.c \right) -\sum_{i,a} (\mu+V/2) n^a_{i}\\&+& \frac{V}{2}
     \sum_{i} (n^a_{i})^2-\frac{1}{\beta}\ln \sinh\left( \beta\Delta \sum_a n^a_{i}\right)\nonumber\\
     &+&\frac{1}{\beta}\ln \left(\beta \sum_a n_i^a
     \right)\nonumber,
\end{eqnarray}
where the last two terms are interactions generated by the integration over the disorder. Note the $\beta\Delta^2$ reduction of the on-site repulsion is absent in the uniform distribution case. Consequently, the $T=0$ limit can be taken
explicitly.
Introducing $\psi_i^a$ and still choosing the basis that diagonalizes $H_0$
to perform the cumulant
expansion, we compute the last two terms at $T=0$
and the $n\rightarrow 0$ limit to be
\begin{eqnarray}
\lim_{\beta \rightarrow +\infty} &-&[ \frac{1}{\beta}\ln \sinh(
\beta \Delta\sum_a m^a_{i}\pm 1)
      - \frac{1}{\beta}\ln (\beta \sum_a  m_i^a  \pm
     1)  ] \nonumber\\
     &+&[ \frac{1}{\beta} \ln \sinh( \beta \Delta \sum_a m^a_{i} )
     - \frac{1}{\beta}\ln (\beta \sum_a  m_i^a  ) ]\nonumber\\
     &=&-\lim_{\beta \rightarrow +\infty}\lim_{n\rightarrow
     0}\frac{1}{\beta}\ln\left[\frac{\sinh(\beta \Delta n m^a_{i}\pm \beta\Delta )}{\sinh(\beta \Delta n
     m^a_{i})}\right]\nonumber\\
     &=& -\lim_{\beta \rightarrow +\infty}\lim_{y=n \beta\Delta m\rightarrow
     0}\frac{1}{\beta}\ln\left[\frac{\sinh(y\pm \beta\Delta )}{y}\right]\nonumber\\
     &=&  -\lim_{\beta \rightarrow +\infty}\frac{1}{\beta}\ln\cosh(\pm \beta\Delta )\nonumber\\
     &=& -\Delta   \nonumber
\end{eqnarray}
Thus we have,
\begin{eqnarray}\label{EUD}
  \varepsilon_+(m)&=&
m V-\mu-\Delta \nonumber\\  \ \varepsilon_-(m)&=&(1-m)
 V+\mu-\Delta \ (\rm uniform)
\end{eqnarray}
where  $\mu$ is replaced by $\mu+\Delta$ in the
first term and by $\mu-\Delta$ in the second term.  It is this
structure that makes the width of the MI lobes shrink by $\Delta$ as a function of filling
relative to that in the clean limit.  We then use Eq. (\ref{pb}) to obtain
\begin{eqnarray}\label{bose}
%\frac{\mu}{V}&=& m-\frac{1}{2}+\frac{J}{2V}\pm\left[ 1-\left(4m+2-4\Delta\right)\frac{J}{V}\right.\nonumber\\
%&+&\left.\left(1+(8m+4)\Delta+4\Delta^2\right)\left(\frac{J}{V}\right)^2
%\right]^{1/2}
y&=& m-\frac{1}{2}-\frac{x}{2}\nonumber\\
&\pm& \frac{1}{2}\sqrt{(1-2 \delta)^2+\left(4m+2\right)(2
\delta-1)x+x^2}
\end{eqnarray}
as the phase boundary in the $x-y (J/V-\mu/V)$ ($x$ and $y$
represent $J/V$ and $\mu/V$ he) plane for the Mott
insulator-superfluid transition. Here $\delta=\Delta/V$.

\begin{figure}
  \includegraphics[scale=0.50]{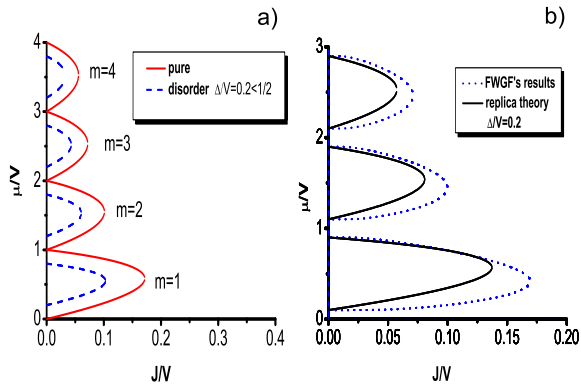}\\
  \caption{a) Phase diagram for disordered Bose-Hubbard model with uniform distribution
  with $\delta=\Delta/V=0.4$. b) Comparison between the replica theory and the treatment of Fisher,
  et al.\cite{fisher} for the infinite-range hopping model with a uniform distribution of site energies. }\label{fig4}
\end{figure}

The phase diagram in the $x-y$ plane shown in Fig. (\ref{fig4}a)
bares close resemblance to the finite temperature counterpart of the
Gaussian distribution.  The only difference between the two is that
the disorder in the Gaussian case is characterized by
$\alpha=\beta\Delta^2$, whereas for the uniform distribution at
$T=0$, the strength of the disorder is set by $\delta=\Delta/V$.
Consequently, in the uniform distribution, the Mott lobes display a
vertical shift of $\delta$ rather than $\alpha$ as in the Gaussian
case. For an independent check on the accuracy of the replica
method, we consider the uniform distribution but with infinite range
hopping.  In Fig. (\ref{fig4}b) we compare the replica method with
the  mean-field criterion
\begin{equation}\label{infinite}
    x=- \frac{2\delta}{\ln\left[\frac{\left(m-(y+\delta)\right)^{(m+1)}
    \left(m-1-(y-\delta)\right)^{m}}{\left(m-(y-\delta)\right)^{(m+1)}\left(m-1-(y+\delta)\right)^{m}}\right]}
\end{equation}
derived by Fisher, et al.\cite{fisher}.  As is evident, only minor quantitative differences obtain, lending credence to the replica treatment presented here.

\subsection{Bose Glass}In the dirty boson model, a localized phase (Bose glass) exists in which disorder rather than the on-site repulsion (Mott insulator) is the root cause.  Unlike traditional spin glass phases which are characterized by an Edwards-Anderson order parameter, the Bose glass does not admit such a description.  In fact for the Bose-Hubbard model, the only Edwards-Anderson parameter that could be non-zero is $\langle b_i^a(t) b_i^c(t')\rangle$. For the superfluid phase, this order parameter is trivially non-zero.  However, there is no phase in which such order exists without simultaneously relying on superfluid order.   With nearest-neighbour Coulomb interactions, such a glass is possible\cite{dobro} independent of superfluidity. The current analysis is limited, however, solely to the on-site Coulomb case.

To analyze the Bose glass, we use the standard\cite{fisher,Grinstein,Giamarchi,Dotsenko} one-loop renormalization group equations in conjunction with the mean-field phase boundaries to derive a criterion for the onset of the Bose glass phase.  The field
theory of our model is,
\begin{eqnarray}\label{}
S(\psi)&=& \sum_{k,a}\left[1-J\int
d\tau\langle T_{\tau}b(\tau)b^{\dag}(0)\rangle\right]|\psi^{a}(k)|^2\nonumber\\
&+&\sum_{k}\frac{(ka_0)^2}{2}|\psi^{a}(k)|^2+\sum_{i,a} g_{aa}|\psi_i^a|^4\nonumber\\
& +& \sum_{i,a\ne b}g_{ab}|\psi_i^a|^2|\psi_i^b|^2+O(|\psi|^6)
%S(\psi)&=& \frac{1}{\beta}\int d\tau \left\{\sum_{i,a}\left[1-J\int
%d\tau<T_{\tau}b(\tau)b^{\dag}(0)>\right]|\psi^{a}|^2\right.\nonumber\\
%&+&\frac{(ka)^2}{2}|\psi_i^{a}|^2+J^2 u\sum_{i,a} |\psi_i^a|^4\nonumber\\
%& +&\left.  J^2 \frac{1}{\beta}\int d\tau' v\sum_{i,a\ne b}|\psi_i^a(\tau)|^2|\psi_i^b(\tau')|^2+O(|\psi|^6)\right\}\nonumber\\
\end{eqnarray}
where $a_0$ is the lattice constant.
%\begin{eqnarray}\label{action}
%S(\psi)&=& \beta \sum_{i,a} r_{ij}\psi_i^{a*}\psi_j^a+c.c+\frac{g_{aa}}{4}\sum_{i,a} |\psi_i^a|^4 \nonumber\\
% &+& \frac{g_{ab}}{4}\sum_{i,a\ne b}|\psi_i^a|^2|\psi_i^b|^2+O(|\psi|^6)\nonumber\\
 %S(\psi)&=&  \int d\tau\sum_{i,a} r_{ij}\psi_i^{a*}(\tau)\psi_j^a(\tau)+\frac{g_{aa}}{4}\int d\tau\sum_{i,a} |\psi_i^a(\tau)|^4\nonumber\\
 %&+& \frac{g_{ab}}{4}\int\int d\tau d\tau'\sum_{i,a\ne b}|\psi_i^a(\tau)|^2|\psi_i^b(\tau')|^2+O(|\psi|^6)\nonumber\\
%\end{eqnarray}
The coefficients $g_{aa}$ and $g_{ab}$ can be calculated using
the cumulant expansion procedure outlined in the Appendix.  For a Gaussian
distribution, these coefficients are given by
\begin{eqnarray}\label{gaa}
g_{ab}&=&
-\frac{J^2\Lambda^2}{12\pi^2}\left[\frac{(m+1)^2}{\varepsilon_+^2\left(\varepsilon_+
+ D/2 \right)}+
\frac{ m^2}{\varepsilon_-^2\left(\varepsilon_- +D/2\right)}\right.\nonumber\\
 &+& \left.\frac{
 m(m+1)}{(V-3D)}\left(\frac{1}{\varepsilon_+}+\frac{1}{\varepsilon_-}\right)^2\right]\\
  g_{aa} &=&-\frac{J^2\Lambda^2}{48\pi^2}\left[\frac{(m+1)(m+2)}{\varepsilon_+^2 \left[(
m+\frac{1}{2})V-(m+2)D-\mu\right]}\right.\nonumber\\
&+& \left.\frac{m(m-1)}{\varepsilon_{-}^2\left[-(
m-\frac{3}{2})V+(m-2)D+\mu\right]}\right].
\end{eqnarray}
%\begin{eqnarray}\label{gaa}
%  g_{aa}&=&
%  \frac{1}{2}J^2 A(m)^2+\frac{7V}{12\beta}J^2 A(m)^4\\
%  g_{ab}&=&
%  \frac{1}{2}J^2 A(m)^2-\frac{7\Delta^2}{12}J^2 A(m)^4,
%\eeq
%where
%\beq
%  J A(m)&=&  J \int_0^{\beta} \int_0^{\beta} d\tau d\tau' \left\langle T {b_{i}^a}^{\dag} (\tau) b_{i}^a
% (\tau')\right\rangle\nonumber\\
%  &=& \frac{(m+1)x }{m -y-(1+m)\alpha}+\frac{m x}{(1-m)+y+(m-1)\alpha}.\nonumber
%\end{eqnarray}

The signature of the disorder-induced localized phase is the
divergence of the coupling constant for the interaction between
different replicas.  To this end, we derive the one-loop
renormalization equations\cite{Dotsenko}
\begin{eqnarray}
  \frac{d g_{aa}}{d \xi} &=& \epsilon  g_{aa}\nonumber\\
  &-&  K_d \left((p+2)g_{aa}^2+p\sum_c g_{ac}g_{ca}\right) \nonumber\\
   \frac{d g_{ab}}{d \xi} &=& \epsilon  g_{ab}+ K_d
 \left( (4+2p)(g_{aa}+g_{bb})g_{ab} \right.  \nonumber\\
 &+&\left. 4 g_{ab}^2+ p\sum_c g_{ac}g_{ca}\right)
   \nonumber
\end{eqnarray}
for the coupling constants $g_{ab}$ and $g_{ab}$.  Here $\xi$ is the
standard rescaling parameter, $\epsilon=4-(d+z)$,
$K_d=\frac{2}{(4\pi)^{d/2}\Gamma(d/2)}$, ($K_2=\frac{1}{2\pi}$),
 $p$ is the number of the component of $\psi$ ( $p=2$ in this case)
 and $d$ is the spatial dimension.  We are particularly
interested in $p=2$ and $d=2$. Care must be taken in analyzing these
equations, however, as the coupling constants, $g_{ab}$ and $g_{aa}$
are actually ultrametric matrices. Using the Parisi\cite{parisi}
multiplication rule for such matrices, we partition $g_{ab}$ into a
diagonal part $\tilde{g}$ and an off-diagonal part which is a
function $g(x)$ defined in the domain $x\in (0,1)$. At the replica
symmetric fixed point, we find that
\begin{eqnarray}\label{FP}
    \tilde{g}&=&\frac{\epsilon p}{16(p-1)K_d}\ \\
        g(x)&=&-\frac{\epsilon (4-p)}{16(p-1)K_d} \ {\rm for}\ x\in[0,1].
\end{eqnarray}
This fixed point is unstable\cite{Grinstein} for $p>4(1-\epsilon)$.
That is, for $d< 3.5$, there is a runaway to the strong disorder
region signalled by  $g(x)\rightarrow \infty$, the signature of
localization. The main criterion for the boundary to separate the
disorder relevant and disorder irrelevant regions comes from the
renormalization equation for $g(x)$. If we consider the replica
symmetric case, we only need two parameters, the off-diagonal,
$g(x)=g$, and diagonal parts, $\tilde{g}$. The renormalization
equations simplify to
\begin{eqnarray}
  \frac{d \tilde{g}}{d \xi} &=& \epsilon  \tilde{g}-
  K_d \left[(p+2)\tilde{g}^2+p g^2\right] \nonumber\\
   \frac{d g}{d \xi} &=& \left[\epsilon-K_d
(4+2p)\tilde{g}\right] g +(4-2 p) g^2.
   \label{grg}
\end{eqnarray}
%==============================================================
% Stability of the disordered fixed point
%===============================================================
% Linearized near this fixed
%point, we can find this disordered fixed point is unstable when
%$\epsilon<0$ and stable when $\epsilon>0$\cite{Grinstein,Dotsenko}.
%This fixed point is always unstable under the replica symmetry
%breaking\cite{Dotsenko}. All these make it not a suitable
%representation of the Bose Glass.

Two characteristic properties of the Bose glass in this RG scheme
are 1) $g\rightarrow\infty$, and 2) $\psi=0$.   As is well known,
when $g\rightarrow +\infty$, the RG procedure breaks down.  Hence,
we can use the RG procedure to demarcate the boundary between the
disorder relevant and disorder irrelevant regimes.  From Eq.
(\ref{grg}), the condition for $g$ to run to infinity is
\begin{equation}\label{gcond1}
\epsilon-K_d (4+2p)\tilde{g}>0.
\end{equation}
So the boundary separating disorder relevant region and disorder irrelevant region is,
\begin{equation}\label{gcond2}
 \epsilon-K_d (4+2p)\tilde{g}=0.
\end{equation}

 This result is in fact similar to the long
wavelength limit derived by Fisher et al\cite{fisher}. In fact, they
applied the replica trick and RG analysis to a similar mean field
action. Without considering the $p-$ dependence, they found that the
coefficient of $g^2$ (see Eq. (\ref{grg})) is always positive.
However, as is clear from Eq. (\ref{grg}), in the general case when
the $p-$dependence is considered, this coefficient can be negative.
Note the presence of $p$ is two-fold as it also generates a cross
term $\tilde g g$ in the RG equations.

Eq. (\ref{gcond2})  together with that for $\psi=0$, that is, $r>0$
give rise to the Bose glass phase boundary in the phase diagram.
This criterion depends on $\epsilon$, $x=J/V$, $y=\mu/V$, disorder
and the momentum cutoff. Hence, if $\epsilon$, the disorder, and the momentum cutoff
are given, Eq.(\ref{gcond2}) will define a series of curves in the $x-y$ plane.
For Gaussian disorder, in the domain $\mu/V \in
(m+\frac{1}{2}-2\delta,m-\frac{3}{2}+\delta)$, $\tilde{g}<0$. As a result, if
$\epsilon>0$, Eq. (\ref{gcond1}) is always
satisfied, which means that for systems with $d+z<4$, disorder is
always relevant. In this case, the $\psi=0$ regions all turn into
the Bose glass and a direct transition between a Mott insulator and the superfluid is not possible.  If $\epsilon<0$, the criterion (Eq. (\ref{gcond1})) will separate
disorder relevant and irrelevant regions in the $x-y$ plane. In general, the
 criterion depends on $x$, $y$, the disorder, $\delta$ and the momentum
cutoff $\Lambda$, and is given by
%=============================================================================
%####WU explain how you derived this equation.
%=============================================================================
\begin{eqnarray}\label{}
 x_d &=&x_d(y,\Lambda,\epsilon)=\sqrt{\frac{-48\pi^2\epsilon}{\Lambda^2/V}}\nonumber\\
 &\times&\left\{
\frac{(m+1)(m+2)}{\left[m-(m+1)\alpha-y\right]^2\left[m+\frac{1}{2}-(m+2)\alpha-y\right]}\right.\nonumber\\
&+&\left.\frac{m(m-1)}{\left[1-m+(m-1)\alpha+y\right]\left[\frac{3}{2}-m+(m-2)\alpha+y\right]}\right\}^{-1/2}.\nonumber
\end{eqnarray}
For different fillings $m$, we have a class of curves which form
concentric lobes $x=x_d(y,\Lambda,\delta)$ in the $x-y$ plane. So for
each filling number $m$ with $\epsilon<0$, we have a critical value
$x_d$ . If $x>x_d$, disorder relevant, while for $x<x_d$
disorder is irrelevant.
%===========================================================================
%#####Wu could you say more about this.  Be careful with your English.
%You need to read what you write out loud to make sure it makes sense
%as written English.
%=======================================================================

% Figures for g_ab curves

\begin{figure}
  \includegraphics[scale=0.80]{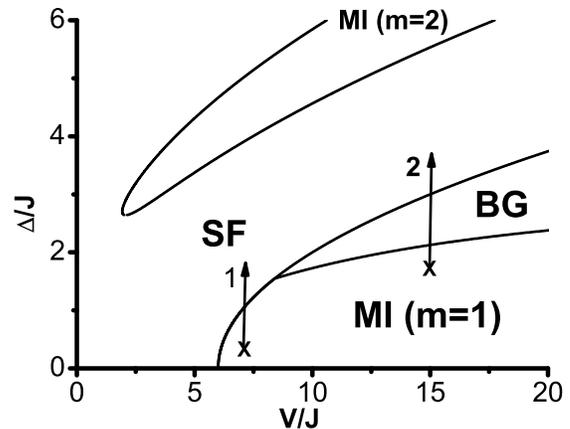}\\
  \caption{A typical phase diagram in $\Delta/J-V/J$ plane for disordered Bose-Hubbard with Gaussian distributed disorder.
  There are three phase: MI,BG and SF. Phase at "$\rm X$" which is a MI
  becomes a SF when the disorder is increased.}\label{fig7}
\end{figure}

The tips of the MI lobes are at
$y=(m+1)\alpha-1+(1-2\alpha)\sqrt{m(m+1)}$, precisely where  $x$
reaches its maximal value.  Recall $\alpha=\beta\Delta^2/V$.  For
large fillings, $m\rightarrow \infty$ and $y$ approach to $y_0$
where $y_0=m-\frac{1}{2}-m\alpha$.
%===========================================================================
%####What is y_0??????? explain
%===========================================================================
To see whether the disorder-relevant region lies within the MI lobes
in $x-y$ plane (recalling that $x=J/V$ and $y=\mu/V$) , we calculate
the value of $x$, $x^{\rm MI}$, for the MI lobes evaluated  at $y_0$ and
$x^{\rm BG}$ of BG lobes evaluated at $y_0$. Consequently, we
consider the ratio,
%=========================================================================
%######Wu it is not clear what x represents...so all of this is quite mysterious.
%###Make it understandable to a group member.
%=========================================================================
$\frac{x^{\rm BG}}{x^{\rm MI}}$. This ratio
\begin{eqnarray}\label{ratio}
    \frac{x^{\rm BG}}{x^{\rm MI}}&=&\frac{2\pi\sqrt{6(-\epsilon)(1-2\alpha)}}{\Lambda^2/V}
    \left(\frac{2m+1}{\sqrt{m^2+m+1}}\right)
\end{eqnarray}
is a decreasing function of filling number $m$, and for large
filling number, \beq\label{ratio2}
  \frac{x^{\rm BG}}{x^{\rm MI}}= \frac{4\pi\sqrt{6(-\epsilon)(1-2\alpha)}}{\Lambda^2/V }, {\rm as} \ m\rightarrow \infty .
\eeq Whether we can have a direct transition from MI to SF depends on
the whether the above ratio is greater or less than one.
If the above ratio is greater than one, it means that the curve
demarcating the disorder-relevant region intersects the $r=0$ curve.
In this case, the BG region is located in the upper and lower regions
of MI lobes as depicted in Fig. (\ref{bg}). Consequently, in such
cases, a direct transition from MI to SF is allowed. If the above
ratio (Eq.(\ref{ratio2})) is less than one,
%====================================================
%####Specify the Equation.
%===================================================
 the BG surrounds the MI and a direct transition
between the MI and SF is forbidden.
 Because for $m=1$, $x_c$ runs to infinity as
$y\rightarrow 0$, the disorder curve always intersects the $r=0$
curves at $m=1$. Consequently, we reach the conclusion that
for $m=1$, a direct transition is always
allowed.  This prediction is in principle testable by direct numerical simulation. The phase diagram, Fig. (\ref{fig7}), in the $\Delta/J-V/J$ plane displays the direct transition from the MI to the superfluid as the disorder is increased. In this plane, a further increase in the disorder leads to a transition to $m=2$ Mott insulating state.  Hence, we predict that the superfluid density should be a non-monotonic function of the disorder. Similar conclusions have been reached in a Landau-Ginzburg treatment\cite{lgdis} of the supersolid problem.

The ratio in Eq. (\ref{ratio}) depends on the momentum cutoff $\Lambda$
and the disorder $\alpha$. We can see that increasing disorder $\alpha$
will decrease the ratio, so if $\Lambda$ is less than a critical
value, no matter what strength the disorder is, the ratio is always
less than one; thus direct transitions are forbidden except for $m=1$.
So for a given momentum cut-off $\Lambda$ and interaction $V$, a
direction transition from MI to SF is forbidden if $\Lambda<\Lambda_c$ where
\begin{eqnarray}\label{Lambda_c}
  \Lambda_c^4 = 96\pi^2(-\epsilon)V^2
\end{eqnarray}
%=====================================================
% ###How did you derive this equation.
%=====================================================
which follows from $\frac{x^{\rm BG}}{x^{\rm MI}} <1$ assuming
$\alpha=0$. In this case, for any disorder strength, a direct
transition is impossible between the MI and the SF at large
fillings.  This corroborates the result derived earlier by
Herbut\cite{herbut} that the MI phase is always surrounded by the BG
in the large-filling limit of the Bose-Hubbard model.

For large momentum cutoff $\Lambda>\Lambda_c$, the ratio
$\frac{x^{\rm BG}}{x^{\rm MI}}$ could be greater or less than one
depending on the disorder strength $\alpha$. A critical value of
$\Delta_c$ exists. Hence, for weak disorder, $\Delta<\Delta_c$, we have a
direct transition for large fillings $\Delta<\Delta_c$ with
\begin{eqnarray}\label{dcond}
\frac{\beta\Delta_c^2}{V} =
1-\frac{1}{96\pi^2(-\epsilon)}\left(\frac{\Lambda^4}{V^2}\right).
\end{eqnarray}
which is derived from $\frac{x^{\rm BG}}{x^{\rm MI}} <1$ assuming
$\Lambda>\Lambda_c$.

%======================================================================
%#####Wu this sentence states a contradiction.  Please clarify.
%======================================================================

%===================================================================
% RG modify r=0 boundary?
%===================================================================
Renormalization also modifies the value of $r$ to
\cite{Dotsenko},
\begin{eqnarray}\label{}
r(\xi)&=& r_0\exp\left\{\left[2-K_d((2+p)\tilde{g}+p
g(x)\right]\xi\right\}\nonumber\\
&=& r_0\exp\left\{-\frac{1}{2}\xi\right\}= r_0 R_c^{-1/2}.
\end{eqnarray}
where we have only considered the replica symmetric case. Here $R_c$
is the correlation length $R_c=\exp(\xi)\propto r_0^{-\nu}\approx
r_0^{-1/2}$, so $r(\xi)\propto r_0^{5/4}$ which means that the
renormalization does not shift the MI-SF phase boundary which occurs
at $r_0=0$.

Ultimately, it is the Bose glass that makes the disordered boson
problem distinct from the disordered electronic Mott insulator.  In
the presence of disorder, the boson lattice adjusts (contracts or
expands) so that the chemical potential remains in the gap.  In the
electron problem, in which the electrons occupy pre-existing lattice
sites, disorder changes the position of the chemical
potential\cite{Sawatzky}.  Consequently, for the boson problem, it is
the nature of the in-gap states that ultimately determines whether the
disordered system is localized or not.  However, as we see here the
criterion is a complicated function of the system parameters.

\subsection{$^3$He impurities}

\begin{figure}
  %  \centering
        \includegraphics[scale=0.60]{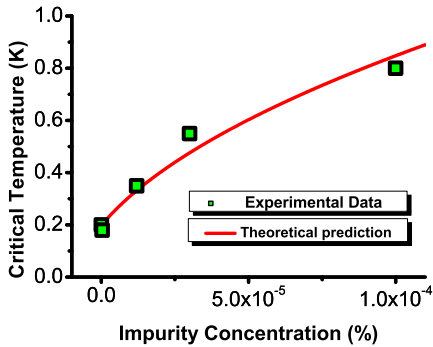}
    \caption{Critical temperature as a function of impurity concentration. Experimental data are taken from Ref. (7).}
    \label{fig1}
\end{figure}

%============================================================
%
%============================================================
$^3$He increase the onset temperature for the missing moment of inertia.
Although we do not have a microscopic model for a grain boundary, the point defect model we have outlayed here explains this effect qualitatively as disorder can enhance the superfluid region.  
In essence, a disordered system with interaction $V$ can be represented by a
pure system with an effective interaction $V_{\rm eff}$. If for a
pure system, the critical interaction is $V_c$, then for a
disordered system, the corresponding critical point is $V_{\rm
eff}=V-\beta\Delta^2=V_c$.  An immediate
consequence is that the new boundary for the Mott-superfluid
transition is shifted to higher values of the on-site interaction.
That is, for the disordered system (with one boson per site) $V_c$
is replaced by 
\beq\label{VCD} V_c(\Delta)=V_c+\beta_c\Delta^2. 
\eeq
 Consequently, to remain on the
phase boundary, increasing the disorder must be compensated by an
increase in the onset temperature as is seen
experimentally\cite{impeff} for $^3$He defects and studied theoretically by a Abrahams and Balatsky\cite{lgdis} using a Landau-Ginzburg approach.  To formalise this,
we consider $^3$He defects with a concentration $c$ and an on-site
energy $\varepsilon_2$. We will treat the $^4$He atoms as having
on-site energy $\varepsilon_1$ with concentration $1-c$. A rigorous
treatment require a binomial distribution of disorder. But to get
the basic scene of the influence of disorder, we use Gaussian
distribution to approach this disorder. 
The key parameter is the variance of the distribution of on-site
energies,
$\Delta^2=\Delta_d^2+c(1-c)(\varepsilon_2-\varepsilon_1)^2$, where
$\Delta_d^2$ is the disorder which can be eliminated by annealing.
 For a
clean system, the transition from the Mott insulator to the
superfluid is given by $k_BT_c/J=(V_c-V)/V_c$\cite{PPbook}.  We now
replace $V$ by $V_{\rm eff}$ and solve for $T_c$. The solution, \beq
    K_B T_c&=& p_1 J+\sqrt{\left(p_1 J \right)^2+p_2 J+p_3 J c\left(1-c\right)},
\eeq has a square-root dependence with $p_1=\frac{(V_c-V)}{2V_c}$,
$p_2=\frac{\Delta_d^2}{V_c}$ and
$p_3=\frac{(\varepsilon_2-\varepsilon_1)^2}{V_c}$. Knowing that the
critical temperature is $0.2 K$ in the absence of $^3$He impurities
leads to a relationship between $p_2 J$ and $p_1 J$. Thus, we have
two free parameters $p_1 J$ and $p_3 J$ to fit the curve. We show in
Fig. (\ref{fig1}) a plot with the fitting parameters: $p_1
J=-0.10 K$, $p_3 J=(90 K)^2$ and $p_2 J= (0.28 K)^2$.  From the
above formula, we can see that if there are no impurities and no other disorder that can be annealed away, that is, both $c=0$ and $\Delta_d=0$, we obtain
a negative $T_c$ which means
there is no supersolid transition. Also, if disorder is too large,
$V_{\rm eff}=V-\beta\Delta^2<0$, and the "net interaction" is
attractive which results in an insulating phase. Consequently, for sufficiently large disorder, we also obtain an absence of a supersolid transition.
Hence, although the treatment here is not rigorous, it sufficiently rich to capture the interplay between disorder, finite temperature, and supersolidity. 
The quantitative agreement, which is tied more to the functional form
than the fitting parameters, lends credence to our claim that
disorder underlies the missing moment of inertia in solid $^4$He.

\section{Conclusion}

We have presented what we think is the minimal model that captures disorder-induced superfluidity in bosonic systems.  While we undoubtedly do not have the sufficient microscopic details to model actual grain boundaries, the results presented here offer a general framework in which the general problem of disorder-induced superfluidity can be formulated consistently.  We have seen from our replica analysis and the one-loop renormalization analysis that the phase boundaries of the disordered Bose-Hubbard model
can be determined but do not appear to be universal, in contrast to the phase boundaries constructed from general considerations in the early work of Fisher, et al.\cite{fisher}.  In particular, a direct
 MI-SF transition is possible as found earlier\cite{bh3,bh5}; however, the criterion depends on the
 disorder, interaction strength and filling numbers.  Further, we have shown how Mott insulating phases can be observed in unbounded distributions.  This application is particularly relevant to experiments\cite{demarco} on optical lattices as the disorder in such systems always obeys an unbounded distribution.  
 %though it vanishes giving rise to the standard BG-SF transition
 %at sufficiently large filling.
 Since the flows are to the strong disorder limit,
a treatment (currently not available) in this parameter space is
essential to understanding the phase structure of the Bose-Hubbard
model. Finally, because the MI phases always give rise to
superfluids for intermediate disorder (for example, $0<D<V$ for
Gaussian distributions), we believe this model is the correct
starting point for analysing the reports of missing moment of
inertia in solid $^4$He induced by disorder, in particular the
extreme sensitivity of the critical temperature to $^3$He
impurities.

\acknowledgements We thank A. Balatsky, for initial discussions which
motivated this work, D. Ceperley for critical advice, D. Huse for several conversations that shaped the final version of this paper and NSF
DMR-0605769 for partial support.

\section{Appendix}
In the section, we will derive the effective action. The action is,
\begin{eqnarray}
S(\psi)&=& \beta \sum_{i,a} r\psi_i^{a*}\psi_j^a+c.c+u\sum_{i,a} |\psi_i^a|^4 \nonumber\\
 &+&  v\sum_{i,a\ne b}|\psi_i^a|^2|\psi_i^b|^2+O(|\psi|^6)
 %S(\psi)&=&  \int d\tau\sum_{i,a} r_{ij}\psi_i^{a*}(\tau)\psi_j^a(\tau)+\frac{g_{aa}}{4}\int d\tau\sum_{i,a} |\psi_i^a(\tau)|^4\nonumber\\
 %&+& \frac{g_{ab}}{4}\int\int d\tau d\tau'\sum_{i,a\ne b}|\psi_i^a(\tau)|^2|\psi_i^b(\tau')|^2+O(|\psi|^6)\nonumber\\
\end{eqnarray}
where in momentum space,
\beq
r=\frac{1}{J\cos(k a_0)}-\int d\tau \langle
T_{\tau}b(\tau)b^{\dag}(0)\rangle
\eeq
 and $a_0$ is the lattice
constant. The coefficients $u$,$v$ are given by the averages,
\begin{eqnarray}
  u &=&-\frac{1}{24}\int d\tau_1...d\tau_4
  \left<T_{\tau}b^a(\tau_1)b^a(\tau_2)b^{a\dag}(\tau_3)b^{a\dag}(\tau_4)\right>\nonumber\\
  &-& \frac{1}{8}\left[\int d\tau_1 d\tau_2\left<T_{\tau}b^a(\tau_1)b^{a\dag}(\tau_2)\right>\right]^2  \\
    v &=&-\frac{1}{24}\int d\tau_1...d\tau_4
  \left<T_{\tau}b^a(\tau_1)b^b(\tau_2)b^{a\dag}(\tau_3)b^{b\dag}(\tau_4)\right>\nonumber\\
  &-& \frac{1}{8}\left[\int d\tau_1 d\tau_2\left<T_{\tau}b^a(\tau_1)b^{a\dag}(\tau_2)\right>\right]\nonumber\\
  &\times& \left[\int d\tau_3 d\tau_4\left<T_{\tau}b^b(\tau_3)b^{b\dag}(\tau_4)\right>\right]
\end{eqnarray}
To compute these correlation functions, we insert a complete set of
states \beq \Pi_{i,a}|m_i^a><m_i^a|=1 \\
\eeq in between all $b^{\pm}(\tau_i)$ operators and integrate over
all $\tau_i$.
%==========================================================================
%####What are you saying here. This sentence is unclear.
%==========================================================================
 The
terms from the first term of the order of $u$, $v$ of which the
order of replica indices  are $aabb$ or $bbaa$ will cancel with the
second term. So we have,
\begin{eqnarray}\label{u}
u &=&
-\frac{\beta}{24}\left[\frac{(m+1)(m+2)}{\epsilon_1^2(\epsilon_1+\epsilon_2)}
+\frac{m(m-1)}{\epsilon_{-1}^2(\epsilon_{-1}+\epsilon_{-2})}\right]\nonumber\\
  v &=& -\beta\sum_{a\neq b}\left[\frac{(m+1)^2}{6\epsilon_{1,0}^2(\epsilon_{1,0}+\epsilon_{2,1})}
  +\frac{m(m+1)}{12\epsilon_{1,0}^2(\epsilon_{1,0}+\epsilon_{0,1})}\right.\nonumber \\
   &+& \frac{m(m+1)}{12\epsilon_{-1,0}(\epsilon_{-1,0}+\epsilon_{0,-1})(\epsilon_{-1,0}+\epsilon_{0,-1}-\epsilon_{0,1})}\nonumber \\
&+& \frac{m(m+1)}{12\epsilon_{1,0}(\epsilon_{1,0}+\epsilon_{0,1})(\epsilon_{1,0}+\epsilon_{0,1}-\epsilon_{0,-1})}\nonumber \\
&+&
\left.\frac{m(m+1)}{12\epsilon_{-1,0}^2(\epsilon_{-1,0}+\epsilon_{0,-1})
}+ \frac{m^2}{6\epsilon_{-1,0}^2(\epsilon_{-1,0}+\epsilon_{-2,-1})
}\right]\nonumber
\end{eqnarray}
where the energies are defined as follows:
\begin{eqnarray}
   E(m_i^a,m_i^b)&=&\frac{V_{\rm eff}}{2}(m_i^{a 2}+m_i^{b 2})-\mu_{\rm
   eff}(m_i^a+m_i^b)\nonumber\\
   &-&\frac{\beta\Delta^2}{2}m_i^a m_i^b\\
   \epsilon_{0,-1}&=& E(m_i^a+1,m_i^b-1)-E(m_i^a,m_i^b-1)\nonumber\\
   &=& m V-(m+2)D-\mu \nonumber\\
 \epsilon_{1,0}&=& E(m_i^a+1,m_i^b)-E(m_i^a,m_i^b)=\varepsilon_+\nonumber\\
 &=& m V-(m+1)D-\mu \nonumber\\
 \epsilon_{2,1}&=& E(m_i^a+1,m_i^b+1)-E(m_i^a,m_i^b+1) \nonumber\\
 &=& m V-m D-\mu\nonumber\\
 \epsilon_{-2,-1}&=& E(m_i^a-1,m_i^b-1)-E(m_i^a,m_i^b-1) \nonumber\\
 &=& -(m-1)V+m D+\mu\nonumber\\
\epsilon_{-1,0}&=& E(m_i^a-1,m_i^b)-E(m_i^a,m_i^b)=\varepsilon_- \nonumber\\
&=& -(m-1)V+(m-1)D\nonumber\\
\epsilon_{0,1}&=& E(m_i^a-1,m_i^b+1)-E(m_i^a,m_i^b+1) \nonumber\\
&=& -(m-1)V+(m-2)D+\mu\nonumber\\
\epsilon_{\pm 1}&=& E(m_i^a\pm 1,m_i^b)-E(m_i^a,m_i^b)=\varepsilon_{\pm} \nonumber\\
 \epsilon_{\pm 2}&=& E(m_i^a\pm 2,m_i^b)-E(m_i^a\pm,m_i^b)\nonumber\\
\epsilon_{-1}+\epsilon_{-2} &=& 2\left[-(
m-\frac{3}{2})V+(m-2)D+\mu\right]\nonumber\\
 \epsilon_{1}+\epsilon_{2} &=& 2\left[(
m+\frac{1}{2})V-(m+2)D-\mu\right].
\end{eqnarray}
Here we have used the trick $\lim_{n\rightarrow 0}\sum_{b\neq a}
A=\lim_{n\rightarrow 0}(n-1)A=-A$. We can simplify the result further to
\begin{eqnarray}
 v &=&\frac{(m+1)^2}{12\varepsilon_+^2\left(\varepsilon_+ + D/2 \right)}
 + \frac{m^2}{12\varepsilon_-^2\left(\varepsilon_- +D/2\right)}\nonumber\\
 &+& \frac{m(m+1)}{12(V-3D)}\left(\frac{1}{\varepsilon_+}+\frac{1}{\varepsilon_-}\right)^2
\end{eqnarray}

To proceed, we make the approximation
\beq
1/[J\cos(k a_0)]\approx1/[J(1-(k a_0)^2/2)]=(1/J)[1+(k a_0)^2/2].
\eeq
We then rescale $\psi$ by the factor
$\psi\rightarrow\sqrt{J}a_0^{d/2-1}\psi$ and replace the $\sum_i$,
$\beta$ by $\int \frac{d^d x}{a_0^d}$, $\int d\beta$ respectively.
We have,
\begin{eqnarray}\label{}
S(\psi)&=& \int d^d x \int
d\beta\left\{\sum_{a}\left[1-J\left(\frac{m+1}{\varepsilon_+}+\frac{m}{\varepsilon_-}\right)\right]\frac{1}{a_0^2}\right.\nonumber\\
&\times&|\psi^{a}(x,\tau)|^2+\frac{1}{2}|\nabla\psi^{a}(x,\tau)|^2+\frac{J^2 u}{a_0^2}\sum_{a} |\psi^a(x,\tau)|^4\nonumber\\
& +&\left.\frac{  J^2 v}{a_0^2} \sum_{a\ne
b}|\psi^a(x,\tau)|^2|\psi^b(x,\tau)|^2+O(|\psi|^6)\right\}
%S(\psi)&=& \frac{1}{\beta}\int d\tau \left\{\sum_{i,a}\left[1-J\int
%d\tau<T_{\tau}b(\tau)b^{\dag}(0)>\right]|\psi^{a}|^2\right.\nonumber\\
%&+&\frac{(ka)^2}{2}|\psi_i^{a}|^2+J^2 u\sum_{i,a} |\psi_i^a|^4\nonumber\\
%& +&\left.  J^2 \frac{1}{\beta}\int d\tau' v\sum_{i,a\ne b}|\psi_i^a(\tau)|^2|\psi_i^b(\tau')|^2+O(|\psi|^6)\right\}\nonumber\\
\end{eqnarray}
Denoting the momentum cutoff $\Lambda=\frac{\pi}{a_0}$, from above
we have,
\begin{eqnarray}\label{}
g_{ab}&=& \frac{J^2}{a_0^2} u=-\frac{J^2\Lambda^2
}{12\pi^2}\left[\frac{(m+1)^2}{\varepsilon_+^2\left(\varepsilon_+ +
D/2 \right)}+
\frac{ m^2}{\varepsilon_-^2\left(\varepsilon_- +D/2\right)}\right.\nonumber\\
 &+& \left.\frac{
 m(m+1)}{(V-3D)}\left(\frac{1}{\varepsilon_+}+\frac{1}{\varepsilon_-}\right)^2\right]\\
  g_{aa} &=&\frac{J^2}{a_0^2}v=-\frac{J^2\Lambda^2}{48\pi^2}\left[\frac{(m+1)(m+2)}{\varepsilon_+^2 \left[(
m+\frac{1}{2})V-(m+2)D-\mu\right]}\right.\nonumber\\
&+& \left.\frac{m(m-1)}{\varepsilon_{-}^2\left[-(
m-\frac{3}{2})V+(m-2)D+\mu\right]}\right]
\end{eqnarray}
where the dependence on the cut-off is allowed.


\begin{thebibliography}{}

\bibitem{vycor} E. Kim and M. H. W. Chan, Nature {\bf 427}, 225 (2004).

\bibitem{KC} E. Kim and M. H. W. Chan, Science {\bf 305}, 1941 (2004).

\bibitem{impeff} A. C.  Clark and M. H. W. Chan, J. Low Temp. Phys. {\bf 138}, 853 (2005).

\bibitem{Chan06} E. Kim and M. H. W. Chan, Phys. Rev. Lett. {\bf 97}, 115302 (2006).

\bibitem{null1}E. Kim and M. H. W. Chan, Phys. Rev. Lett. {\bf 97}, 115302 (2006).

\bibitem{null2}A. Penzev, et al., cond-mat/0702632.

\bibitem{null3}K. Shirahama, et al., Bull. Am. Phys. Soc. {\bf 51}, 450 (2006).

\bibitem{SRR} A. S. C. Rittner and J. D. Reppy, Phys. Rev. Lett. {\bf 97}, 165301 (2006).

\bibitem{rittner}A. S. C. Rittner and J. D. Reppy, Phys. Rev. Lett. {\bf 98}, 175302 (2007).

\bibitem{SB} S. Sasaki, et al. Science {\bf 313}, 1098 (2006).
%===============================10======================
\bibitem{g1}M. Kubota, et al., unpublished (2006).

\bibitem{g2}K. Shirahama, et al., unpublished (2006).


\bibitem{science}For a review, see P. Phillips and A. V. Balatsky, Science {\bf 316}, 1435 (2007).

\bibitem{leggett} A. J. Leggett, Phys. Rev. Lett. {\bf 25}, 1543 (1970).

\bibitem{chester67} G. V. Chester and L. Reatto, Phys. Rev. {\bf 155}, 88 (1967).

\bibitem{channew}A. C. Clark, J. T. West, and M. H. W. Chan,
cond-mat/07060906.

\bibitem{todo} I. A. Todoshchenko , H. Alles, H. J. Junes, A. Ya. Parshin,
and V. Tsepelin, cond-mat/0703743.

\bibitem{AL} A. F. Andreev and I. M. Lifshitz, Sol. Phys. JETP{\bf 69}, 1107 (1969).

\bibitem{anderson}P. W. Anderson, W. F. Brinkman, and D. A. Huse, Science {\bf 310}, 1164(2005).


\bibitem{avb}Z. Nussinov, A.V. Balatsky, M.J. Graf, and S.A. Trugman, cond-mat/0610743,
submitted to Phys. Rev. B; A.V. Balatsky, Z. Nussinov,  M. Graf and
S. Trugman, Phys. Rev. {\bf B 75}, 094201, (2007).

\bibitem{beamish}J. Day and J. Beamish, Phys. Rev. Lett. {\bf 96}, 105304,
(2006).
%================================20=========================================
\bibitem{gb1} E. Burovski, et al., Phys. Rev. Lett. {\bf 94}, 165301 (2005); N. Profkof'ev and B. Svistunov, Phys. Rev. Lett. {\bf 94}, 155302 (2005).

\bibitem{CC} B. K. Clark and D. M. Ceperley, Phys. Rev. Lett. {\bf 93}, 155303 (2004).

\bibitem{huse}D. Huse, unpublished.


\bibitem{Sawatzky} S. Yunoki and G. A. Sawatzky, cond-mat/0110602.

\bibitem{vac}M. Boninsegni, Phys. Rev. Lett. {\bf 97}, 80401 (2006).

\bibitem{fisher}M. P. A. Fisher,  et al., Phys. Rev. B {\bf 40}, 546 (1989).

\bibitem{scall}R. T. Scalettar, G. G. Batrouni, and G. T. Zimanyi, Phys. Rev. Lett. {\bf 66}, 3144 (1991); M. Wallin, E. Sorenson, S. Girvin, and A. P. Young, Phys. Rev. B {\bf 49}, 12115 (1994).

\bibitem{bh1} W. Krauth, N. Trivedi, and D. Ceperley, Phys. Rev. Lett. {\bf 67}, 2307 (1991); K. Singh and D. S.Rokhsar, Phys. Rev. B {\bf 46}, 3002 (1992); P. Sen, N. Trivedi, D. M. Ceperley, Phys. Rev. Lett. {\bf 86}, 4092 (2001); M. Makivic, N. Trivedi, S. Ullah, ibid {\bf 71}, 2307 (1993).

\bibitem{bh2}J. Freericks and H. Monien, Phys. Rev. B {\bf 53}, 2691 (1996).
%==========================30==========================================
\bibitem{bh3}J. Kisker and H.  Rieger, PRB, R11981 vol. 55 (1997).

\bibitem{bh4}R. V. Pai, et al. Phys. Rev. Lett. {\bf 76}, 2937 (1996);
N. Prokof'ev and B. V. Svistunov, Phys. Rev. Lett. {\bf 80}, 4355 (1998);
B. V. Svistunov, Phys. Rev. B {\bf 54}, 16131 (1996).

\bibitem{bh5} F. Pazmandi and G. T. Zimanyi, Phys. Rev. B {\bf 57}, 5044 (1998).

\bibitem{herbut}I. Herbut, Phys. Rev. Lett. {\bf 79}, 3502 (1997); ibid Phys. Rev. B {\bf 57}, 13729 (1998).

\bibitem{bh7}M. B. Hastings, Phys. Rev. B {\bf 64}, 24517 (2001).

\bibitem{demarco}B. DeMarco, private communication.  In his experiments, the distribution characterising the disorder decays as an exponential rather than Gaussian.  We have also analysed this case and find that our conclusions hold as well.

%\bibitem{simmons}S. M. Heald, D. R. Baer, and R. O. Simmons, Phys. Rev. B {\bf 30}, 2531 (1984).

%\bibitem{gbound} R. Roth and K. Burnett, Phys. Rev. A {\bf 68}, 23604 (2003).


%==========================================================================
\bibitem{PPbook} P. Phillips, Advanced Solid State Physics, (Westview, Colorado) (2003).

\bibitem{RSY} N. Read, S. Sachdev and J. Ye, Phys. Rev. B {\bf 52}, 384 (1995).

%\bibitem{huse2}
%J. Miller and D. A. Huse, Phys. Rev. Lett. {\bf 70}, 3147 (1993); A.
%J. Bray and M. A. Moore, Phys. Rev. Lett. {\bf 13}, L655 (1980).

\bibitem{WP} J. Wu and P. Phillips, Phys. Rev. B {\bf 73}, 214507 (2006).

\bibitem{DP} D. Dalidovich and P. Phillips, Phys. Rev. B {\bf 59}, 11925

(1999).

\bibitem{DP2}D. Dalidovich and P. Phillips, Phys. Rev. Lett. {\bf 89}, 27001 (2002).

\bibitem{DP3} P. Phillips and D. Dalidovich, Phys. Rev. B {\bf 68}, 104427 (2003).

%\bibitem{DD} D. Dalidovich and V. Dobrosavljevi$\acute{c}$ , Phys. Rev. B {\bf 66}, 081107(2002)

%\bibitem{RR} A. S. C. Rittner and J.D.Reppy,cond-mat/0702665

\bibitem{dobro}S. Pankov and V. Dobrosavljevic, Phys. Rev. Lett. {\bf 94}, 046402 (2005).

\bibitem{Grinstein}G. Grinstein and A. Luther, Phys. Rev. B {\bf 13}, 1329 (1976).

\bibitem{Giamarchi}T. Giamarchi, et al., Phys.Rev. B {\bf 37}, 325 (1988)

%\bibitem{Wick} Wick's theorem can only be applied when the
%Hamiltonian is quadratic forms such as $b_i^{\dag}b_i$. But if we
%choose $H_0=-\mu\sum_{i,a} n_i^a$ and use the cumulant expansion, we
%need to assume $J,V\ll \mu$ which contradict to our physical
%assumption where disorder and interaction are dominent terms. So
%here we involved the interaction term in $H_0$ as Eq.(\ref{H}) at
%the same time do the cumulant expansion on both tunneling term and
%interaction. So the formulae for the coefficients of
%$|\psi^a|^2|\psi^b|^2$ and $|\psi|^4$ are not exactly.

\bibitem{Dotsenko} V. Dotsenko and D. E. Feldman, J.Phys.A: Math. Gen. {\bf 28},5183-5206 (1995)

\bibitem{parisi}M. Mezard and G. Parisi, J. Phys. I {\bf 1}, 809 (1991).

\bibitem{rare}Note the renormalization group equations provide a phase boundary that intersects each lobe at $m\pm\delta$.  Hence, a shortcoming of the current analysis is its inability to say anything definitive about the regions on the y-axis spanning $y\in [m+\delta,m+1-\delta]$.  Such regions are governed by Griffiths singularities and appear to be beyond the scope of our analysis.

\bibitem{mott} N. F. Mott, {\it Metal-Insulator Transitions} (Taylor \& Francis, London, 1974).

\bibitem{lgdis}A. V. Balatsky and E. Abrahams, J. Sup. and Novel Mag. {\bf 19}, 395 (2006).
\end{thebibliography}
\end{document}